%% file: Main.tex
\documentclass[10pt,a4paper]{scrartcl} 
\setkomafont{author}{\sffamily}
\usepackage[utf8]{inputenc}
\usepackage[inline,shortlabels]{enumitem}
\usepackage{siunitx,booktabs}
\usepackage[hidelinks]{hyperref}
\usepackage{eurosym,multirow,colortbl}
\usepackage[dvipsnames]{xcolor}
\usepackage[backend=biber,      
	style=authoryear-ibid,
    	maxcitenames = 2,
    	maxbibnames = 3,
    	doi=false,                  
    	url=false,                  
    	isbn=false,                 
    	natbib=true,
]{biblatex}

\usepackage{booktabs}
\usepackage{graphicx}
\usepackage[normalem]{ulem}
\usepackage{amsmath}
\usepackage[nameinlink]{cleveref}
\setcapindent{1pt}
\title{Countershading coloration in blue shark skin emerges from hierarchically organized and spatially tuned photonic architectures inside skin denticles}
\author{
Viktoriia Kamska$^{1,*}$,
Emeline Raguin$^{2}$,
Bodo D. Wilts$^{3}$,
Luca Bertinetti$^{4}$,
Chiara Micheletti$^{2}$,\\
Clemens Schmitt$^{2}$,
Shahrouz Amini$^{2}$,
Maria Murace$^{2}$,
Frederik H. Mollen$^{5}$,\\
Michael Blumer$^{6}$,
Maite Erauskin Extramiana$^{7}$,
Ruien Hu$^{8}$,
Stefan Redl$^{6}$,
Mason N. Dean$^{1,*}$
}
\date{}
\usepackage{graphicx}
\usepackage{caption}
\usepackage{float}
\usepackage[normalem]{ulem}

\begin{document}
\maketitle

\begin{center}
$^{1}$City University of Hong Kong, Kowloon, Yau Tsim Mong, Hong Kong\\
$^{2}$Max Planck Institute of Colloids and Interfaces, Potsdam, Germany\\
$^{3}$University of Salzburg, Salzburg, Austria\\
$^{4}$B CUBE – Center for Molecular Bioengineering, Dresden, Germany\\
$^{5}$Elasmobranch Research Belgium (ERB), Bonheiden, Antwerpen, Belgium\\
$^{6}$Medical University Innsbruck, Innsbruck, Austria\\
$^{7}$AZTI, Basque Research and Technology Alliance (BRTA), Pasaia Antxo, Spain
$^{8}$Hong Kong Polytechnic University, Kowloon, Hung Hom, Hong Kong
\end{center}

\vspace{0.3cm}

\noindent $^{*}$Correspondence: \{phvkamska, mndean\}@cityu.edu.hk

\vspace{0.5cm}

\newcommand{\md}[1]{\textcolor{blue}{#1}}
\newcommand{\kv}[1]{\textcolor{magenta}{#1}}

\input{1-Introduction}

\input{2.1-Results_and_Discussion}

\input{3.Conclusion}
\input{4.Experimental_Section}
\printbibliography
\end{document}

%% file: 1-Introduction.tex
\begin{abstract}
The blue shark (\textit{Prionace glauca}) exhibits a striking dorsoventral color gradient, transitioning from vibrant blue dorsally to silver and white ventrally—a pattern widely interpreted as pelagic countershading. Despite its ecological significance, the physical basis of this coloration remains unresolved. Here we show that this color system does not arise from dermal chromatophores, as in most vertebrates, but from a previously unrecognised photonic architecture housed within the pulp cavity of individual dermal denticles that cover the skin. Optical imaging reveals discrete color domains within denticle crowns, while external denticle morphology remains similar across color zones. Using spectroscopy, micro-computed tomography, histology and correlative electron microscopy, we demonstrate that color variation is organized across coupled micro- and nanoscale architectures. In blue denticles, iridophores and melanophores form a densely packed tessellated reflector–absorber system within an expanded crown-restricted pulp cavity. Transition-zone denticles exhibit partial cellular layering, whereas white denticles lack melanophores and contain only reflective cells. At the nanoscale, ordered purine-crystal stacks generate narrowband blue reflection, whereas disordered assemblies produce broadband white scattering. Together, these results reveal denticles as mechanically protected optical “pixels” whose hierarchical cellular and nanocrystal organization generates the shark’s countershaded coloration.
\end{abstract}
\section*{Introduction}
Skin coloration in pelagic fishes arises from interactions between ambient light, the optical properties of color-producing tissues, and the visual ecology of potential observers \citep{stevens2011animal, mcfall1990crypsis, marshall2003visual}. In open-ocean habitats, where long wavelengths are rapidly attenuated with depth, body colors that mimic the blue–green photons dominating the underwater light field can have a camouflaging effect, influencing predator stealth and prey longevity \citep{stevens2011animal, lowe1996suntanning}. Blue in vertebrate skin is almost always the result of structural color, arising through wavelength-selective scattering from tissue nanostructures rather than purely from pigmentary absorption \citep{fujii1993cytophysiology} (one lineage of teleost fishes with blue pigment being the sole exception; \citep{bagnara2007blue}). The only known exception among vertebrates —a blue pigment cell type known as the cyanophore— has been reported in a single teleost lineage. Whereas most blue structural colors have been examined in terrestrial ecosystems where the color is conspicuous, studying blue-green skins in open ocean swimmers affords an opportunity to understand tissue architectures likely evolved for concealment.
Although the blue shark (\textit{Prionace glauca}) is celebrated for the striking color of its dorsal body skin, it remains unknown whether this optical appearance relies on structural colors or is purely pigmentary. In fact, the mechanisms underlying the optical appearance of shark skin have barely received any focused attention, with the exception of studies on the photophores of deep‑sea bioluminescent lanternsharks \citep{claes2013deepwater,claes2014iso,duchatelet2021third,duchatelet2020extraocular,duchatelet2019etmopteridae}, UV‑induced “suntanning” (melanin increase) in hammerhead sharks \citep{lowe1996suntanning}, and rare pigmentary anomalies such as leucism \citep{Wargat2023Leucism} and xanthism in nurse sharks \citep{macias2025first}. Moreover, it is counterintuitive how sharks can exhibit such a diversity of colors and variegated patterns \citep{dahl2019sandy, skelton2024observations, Whitehead2025}, when their bodies are sheathed in an armor of denticles, toothlike scales jutting up from and often occluding view of the skin \citep{gabler2021dermal, epstein2025zooming}. 

In this study, we investigate the physical basis of the blue shark’s dorsoventral coloration by examining the optical architecture of blue shark skin across multiple spatial scales. Because dermal denticles form the outermost interface between the animal and its optical environment, we test the hypothesis that these structures contribute directly to skin coloration rather than merely overlying pigmentary tissues in the dermis. To evaluate this, we combine optical imaging, reflectance spectroscopy, micro-computed tomography, histology, and electron microscopy to characterize both the external morphology of denticles and the internal organization of their pulp cavities. By linking spectral measurements with cellular and nanoscale structural analyses, we assess how reflective and absorptive components within denticles interact to shape the spectral and spatial properties of reflected light across the shark’s body surface.

%% file: 2.1-Results_and_Discussion.tex
\section*{Result and Discussion}
\subsection*{Correlated Denticle Coloration and Spectral Reflectance Profiles}
\input{Fig1}
To understand how the blue shark’s visible colors emerge, we analysed long swaths of fresh skin, grading from blue dorsally into white on the animal’s belly, countershading widely interpreted as an optical strategy for minimizing body contrast in heterogeneous light fields \citep{stevens2011animal, johnsen2001hidden, rowland2009abbott, stevens2011animal}. From our low-magnification optical imaging, this color gradient contains three regions that visually transition into one another: a blue dorsal zone (BZ), a silver transition zone (TZ) along the flanks, and a white ventral zone (WZ) (\cref{fig1_color_zones}a). From the distinctness of these separate colors, one might reasonably expect the underlying color gradient in blue sharks to arise from distinct changes in denticle morphology, since these should be the first structures contacted by incident light. However, stereomicroscopy and micro-computed tomography ($\mu$CT; 2.5 $\mu$m) showed no differences in the external morphology of denticles among the three regions (\cref{fig1_color_zones}b, left), suggesting that the gross external shape of denticles does not account for the observed color variation. 
Instead, mid-magnification stereomicroscopy revealed that the color of different macroscopic skin regions arises from discrete oval color patches visible within individual denticles (\cref{fig1_color_zones}b, right). Through microdissection and isolation of individual denticles, we verified that these color patches are confined to the crown of each denticle, rather than in the dermis beneath them (\cref{fig2_Structural_Comp} c). To our knowledge, shark denticles have never been reported to house the body’s color: although shark denticle morphology has a long history of study (predominantly to determine links to ecology and locomotion; e.g. \citep{ankhelyi2018diversity, oeffner2012hydrodynamic, rr2024characterization, domel2018shark}, denticles have predominantly been imaged by scanning electron microscopy, due to their diminutive size ($\sim$ \SIrange{150}{300}{\micro\meter}) \citep{rangel2017preliminary, dingerkus1986application, poscai2017microscopic, popp2020denticle}. We believe this has led many works to overlook the skin's surface optical appearance, potentially missing a critically important aspect of how sharks are perceived in their habitats.
 \\
Importantly, the color patches within the denticles of the different skin regions of the blue shark have visibly different colors and degrees of color homogeneity. Whereas BZ and WZ denticles exhibited saturated patches of blue and white, respectively, when viewed with mid-magnification stereomicroscopy, TZ denticles contained a heterogeneous mixture of blue, silver-white, and gold patches, sometimes even within the same denticle (\cref{fig1_color_zones}b, middle-right). These denticle-level visual differences are supported by reflectance spectroscopy measurements conducted on fresh tissue, recording spectra from local skin regions containing multiple denticles (\cref{fig1_color_zones}c). The BZ exhibited a relatively flat reflectance plateau between \num{280} and \SI{450}{nm}, with minimal angular dependence. 
Such non-iridescent reflectance could involve pigmentary absorption, but the absence of sharp absorption features indicates that pigments alone cannot account for the observed spectrum, suggesting an additional contribution from structural coloration generated by sub-micron tissue organization \citep{kinoshita2008physics, surapaneni2024ribbontail, shawkey2017interactions}. The TZ exhibited multi-peaked spectra, including distinct maxima at \SI{350}{nm} and \SI{450}{nm}, followed by broadband reflectance into the red. These spectra are consistent with contributions from structures with more than one characteristic length scale and/or mixed pigmentary and structural effects, in line with observations from other structurally colored tissues \citep{wilts2015spectrally, grether2004individual}. The WZ showed high-intensity broadband reflectance from \SI{280}{nm} to \SI{800}{nm}, similar to other bright white biological tissues containing dense and disordered arrays of high-index scatterers \cite{syurik2018bio, burresi2014bright}.\\
Together, these observations of consistent mesoscale appearance and spectral measurements across macroscopic regions show that the blue shark’s dorsal–ventral countershading arises from the spatial distribution of denticle-level color patches. Furthermore, our spectra suggest that this countershaded appearance arises from variation in fine-scale structural features within denticles, rather than from changes in pigmentation pattern and density in the underlying dermis, which have been the primary focus of previous studies of shark coloration \citep{Wargat2023Leucism, lowe1996suntanning}. We explore the relationship between tissue architecture and optical properties in the following sections.

\subsection*{Structural Coloration from Nanoscale Components within denticle pulp cavities}

Individual blue shark denticles, like those of other sharks, are small, tooth-like, with wide bases anchored in the dermis, tapering through a narrow neck into a flattened crown terminating in a caudally pointing cusp (\cref{fig2_Structural_Comp}). Like teeth, denticles contain an internal pulp cavity that houses nerves, blood vessels, connective tissue, and odontoblasts extending from the underlying dermis. Although denticle shape is known to vary by species and body region \citep{ankhelyi2018diversity, gabler2021dermal, fath2024patterns}, there is no information on how the interior of denticles varies. In blue shark denticles, this pulp cavity extends deeply into the crown, positioning its contents directly beneath the apical denticle surface where they could potentially influence optical appearance (\cref{fig2_Structural_Comp}).
\input{Fig2}

To investigate the origin of the denticle-embedded color domains described above, we carefully trimmed individual denticle crowns from their necks under a stereomicroscope, enabling access to the basal opening of the pulp cavity. In blue-zone (BZ) denticles, optical imaging of the exposed pulp cavities (\cref{fig2_Structural_Comp}c–e) consistently revealed two visually distinct domains within the pulp material: (i) dark brown blotches (tens of \textmu m in diameter) interspersed with (ii) brighter blue shimmering regions. We verified that this internal material is responsible for the color patches observed in intact denticles \cref{fig1_color_zones} by subjecting trimmed crowns to centrifugation and an ultrasonic bath to dislodge the pulp contents. After this treatment, the denticles were visibly empty, their walls remarkably transparent (\cref{fig2_Structural_Comp}d), indicating that the visible apical coloration arises primarily from components contained within the pulp cavity from color-producing material deposited within the mineralized matrix of the denticle crown itself.

The visually distinct color domains within BZ denticles provide a rare opportunity to directly relate microscale optical appearance to underlying nanoscale tissue components. To visualize these structures at higher resolution, we performed environmental scanning electron microscopy (ESEM) by imaging the pulp cavity from the basal side under hydrated conditions. At intermediate (~40\%) relative humidity, two distinct classes of nanoscale structures become apparent, forming clusters corresponding to the optical domains observed in light microscopy (\cref{fig2_Structural_Comp}f–g): (i) spheroidal bodies occupy the regions that appear dark brown in optical images, whereas (ii) platelet-like structures correspond to the blue shimmering domains. Both structures fall within the submicron size range, although the platelets often exceed \SI{1}{\micro\meter} in planar dimension.

The spheroidal bodies closely resemble melanosomes—pigment-containing, light-absorbing organelles ubiquitous across animal taxa in a wide range of tissues \citep{d2019melanosomes, pinheiro2019chemical}. Their morphology and spatial association with optically dark domains strongly support their identification as melanosomes. Melanin-containing chromatophores are well documented in shark skin and other elasmobranch tissues, where they contribute to pigmentation and photoprotection \citep{Wargat2023Leucism, lowe1996suntanning}. Although morphology alone does not uniquely resolve melanin chemistry, the size, shape, and clustering behaviour of the spheroidal bodies observed here are consistent with melanosomes described in other vertebrate pigment cells \citep{simon2010red, sutter2025multiscale}.

In contrast, the platelet-like structures associated with the blue shimmering regions resemble the high-refractive-index purine crystals found in iridophores of many vertebrate structural color systems \citep{hirsch2017biologically, wagner2023macromolecular, rothkegel2025purine}. Although morphology alone does not permit definitive identification of their chemical composition, the size, shape, and optical association of the platelets observed in blue shark denticles are consistent with this class of biological reflectors \citep{wagner2023macromolecular, pinsk2022biogenic, dearden2018sparkling, gur2024physical}.

Together, these observations indicate that denticle coloration arises from nanoscale components housed within the pulp cavity rather than from pigments embedded in the mineralized denticle wall. The co-localisation of absorptive melanosomes and reflective platelet-like structures suggests a functional combination of pigmentary and structural color elements. In many vertebrate color systems, interactions between absorptive melanophores and reflective iridophores play a key role in shaping the final optical output \citep{vogel2015color, thayer2020structural}. In the following section, we therefore examine how these cellular components are spatially organized within denticle pulp cavities across different skin color zones.

\subsection*{Micro- and nanoscale organization of denticle color-producing components}
\input{Fig3}
Given the localized and homogeneous packing of melanosomes and reflective purine crystal in denticles with distinct optical appearance, we investigated how these color-producing components are organized within denticles. Using light microscopy, histology, micro-computed tomography, and electron microscopy, we examined both the microscale arrangement of chromatophore cells (melanophores and iridophores) and the nanoscale organization of reflective purine crystals within iridophores across denticles from different skin color zones (BZ, TZ, WZ) (\cref{fig3_BZ,fig4_TZ_WZ}). Histological sections show that both melanosome clusters and reflective platelets occur within membrane-bound organelles, indicating that they represent intracellular components of melanophores and iridophores, respectively. This cellular context allows the spatial relationships between absorptive and reflective chromatophore types to be compared across denticle zones. Quantitative micro-computed tomography of individual denticles together with sagittal histological sections revealed systematic differences among zones in both pulp cavity size and chromatophore arrangement. In blue-zone (BZ) denticles, the pulp cavity occupies approximately \SI{25}{\%} of the denticle volume and is densely filled with both iridophores and melanophores arranged in a heterogeneous tessellated pattern (\cref{fig3_BZ}b). In transition-zone (TZ) denticles, the cavity volume is slightly smaller (\SI{22.2}{\%}), and the two chromatophore types form a simple layered configuration in which iridophores overlie melanophores along the apico-basal axis (\cref{fig4_TZ_WZ}b). In white-zone (WZ) denticles, the pulp cavity is smallest (\SI{17}{\%}) and contains only iridophores; melanophores are absent (\cref{fig4_TZ_WZ}g). Structural color systems are highly sensitive to the spatial organization of their optical components \citep{schertel2019structural, vogel2015color, kinoshita2008physics, thayer2020structural, henze2019pterin}. The systematic differences in chromatophore composition and spatial arrangement observed across denticle zones therefore suggest that regional optical properties may arise primarily from how reflective iridophores and absorptive melanophores are organized within the pulp cavity.

High-magnification optical imaging of blue-zone (BZ) pulp cavities reveals distinct nanoscale optical signatures within the two structural cell types (\cref{fig3_BZ}d). Iridophores exhibit bright reflective domains aligned in parallel arrays, whereas melanophores—visible as dark-brown cells in sagittal sections (\cref{fig3_BZ}b,c)—display fine punctate shimmering. These contrasting optical signatures correspond to the ultrastructure observed in electron microscopy, namely stacked plate-like crystals within iridophores and densely packed melanosomes within melanophores.
Environmental scanning electron microscopy shows that BZ melanophores are filled with rounded to spheroidal melanosomes whose surfaces bear nanoscale protruding granules (~\SIrange{10}{60}{nm}; \cref{fig3_BZ}g). Transmission electron microscopy further reveals electron-dense intraluminal inclusions within individual melanosomes (\cref{fig3_BZ}h), which spatially coincide with the shimmering domains observed under combined polarized and bright-field illumination. Sub-granular architectures of this scale have only rarely been reported in vertebrate melanosomes, although hierarchical organization into tens-of-nanometre melanin-rich building blocks is a recurring theme in melanin biology \citep{gorniak2014nano,xiao2018elucidation,camacho2019structural,sutter2025multiscale}. The morphology and optical appearance of these features suggest that they may represent optically distinct subdomains within melanosomes. Because their composition and optical properties remain unresolved, their detailed characterization will be explored in future work.
Focused ion beam–scanning electron microscopy (FIB–SEM) imaging revealed that iridophores across skin zones share the same basic photonic components—plate-like purine crystals—but differ in the morphology, orientation, and spatial coherence of crystal stacks within cells.

In blue-zone (BZ) denticles, iridophores contain extended stacks of plate-like purine crystals that occupy much of the cell interior and form closely spaced, quasi-parallel arrays oriented dorsoventrally within each cell (\cref{fig3_BZ}e). Consistent with the chromatophore arrangement visible in sagittal histological sections (\cref{fig3_BZ}b,c), iridophores are positioned above melanophores relative to the denticle crown surface. In transition-zone (TZ) denticles, iridophores also contain stacked purine crystals (\cref{fig4_TZ_WZ}d), but these stacks are substantially shorter. Crystals remain locally aligned within individual stacks, whereas adjacent stacks are frequently rotated relative to one another and rarely extend across the full height of the cell. In white-zone (WZ) denticles, iridophores lack extended crystal stacks altogether. Instead, plate-like purine crystals occur in irregular aggregates without a consistent stacking axis (\cref{fig4_TZ_WZ}j). Compared with the ordered arrays in BZ denticles and the short stacks in TZ denticles, these crystals exhibit substantially greater variation in orientation and spatial organization. Together, these observations reveal a progressive loss of stack length and orientational coherence from BZ to TZ to WZ denticles.\\
\input{Fig4}

To quantify these nanoscale differences, three-dimensional segmentations of iridophores were used to measure crystal morphology and orientation in the two optically most divergent denticle types, BZ and WZ (\cref{fig4_morph,fig4_angle}). Crystal morphospace analysis revealed broadly overlapping aspect-ratio distributions in both zones (\cref{fig4_morph}c), spanning near-equant to elongated plate-like forms. Crystal habit in biogenic purine assemblies can vary with chemical composition; for example, in zebrafish iridophores the guanine–hypoxanthine ratio genetically controls crystal aspect ratio, and more generally biogenic purine crystals often form molecular alloys whose morphology reflects regulated purine availability in vivo \citep{rothkegel2025purine, deis2025genetic, pinsk2022biogenic, wagner2024rationalizing}. The similar aspect-ratio distributions observed in BZ and WZ denticles therefore indicate that crystal morphology alone does not determine denticle color.
\input{Fig4_morph}

Crystal orientation differs markedly between zones (\cref{fig4_angle}). In BZ denticles, most crystals lie within approximately ±15° of the denticle surface plane and form quasi-parallel stacks, indicating a high degree of orientational order at both the crystal and cellular levels. Comparable alignment of reflective platelets occurs in teleost iridophores, where more than \SI{95}{\%} of crystals are oriented parallel to the reflecting surface to produce efficient multilayer reflection \citep{denton1970review, funt2017koi,jordan2014disordered}. In contrast, WZ denticles exhibit substantially broader angular distributions, with rotations frequently exceeding ±60° along multiple axes and no dominant stacking orientation.
The architectures observed across denticle zones correspond to well-established structure–color relationships in biological photonic systems. In multilayer reflectors composed of stacked guanine crystals, the spectral selectivity and angular properties of reflected light depend strongly on the coherence and orientation of crystal stacks \citep{vogel2015color, thayer2020structural}. Highly ordered stacks of parallel platelets produce spectrally selective reflections, whereas increasing disorder in crystal spacing or orientation broadens reflectance spectra and reduces angular color shifts \citep{jordan2014disordered, wilts2018evolutionary}.
\\
\input{Fig4_angle}

The ordered crystal stacks observed in BZ iridophores are consistent with such multilayer reflectors. In many biological systems, including fish scales and cephalopod chromatophores, reflective iridophores are vertically coupled with underlying melanophores that absorb non-reflected light, thereby increasing color saturation and reducing stray scattering \citep{vogel2015color, thayer2020structural, Amiri2012Chromatophores, Zhao2022Ultrastructure, blumer2024intermediate}. The vertical stratification of iridophores above melanophores in BZ denticles therefore likely enhances the depth and spectral purity of the observed blue coloration.

In TZ denticles, iridophores retain locally ordered crystal stacks but are organized in a simple layered cellular arrangement that reduces direct coupling between reflectors and absorbers. Shorter stacks and increased variation in their orientation are expected to broaden reflectance peaks and reduce spectral selectivity, consistent with structure–color relationships described for guanine multilayers and other biological photonic structures \citep{ funt2017koi, gur2024physical, jordan2014disordered}.
In WZ denticles, the absence of melanophores and the highly disordered arrangement of purine crystals shift the optical regime toward broadband scattering. Comparable disordered guanine arrays occur in teleost iridophores that produce bright silvery or white reflections by scattering light over a wide range of viewing angles \citep{pinsk2022biogenic, gur2017light, jordan2012non, wagner2024structure}. By analogy, the disordered crystal assemblies observed in shark WZ denticles likely act as broadband reflectors that generate structural white.

Together, these observations indicate that denticle coloration arises from hierarchical differences in the organization of shared photonic components. BZ denticles combine extended crystal stacks with vertically coupled absorptive melanophores, TZ denticles retain shorter stacks within a simpler layered cellular architecture, and WZ denticles contain only iridophores with highly disordered crystal aggregates. These progressive changes in reflector organization and absorber coupling provide a structural basis for the transition from directional blue reflection to silvery and ultimately broadband white appearance across the shark body surface.

\subsection*{Optical response of iridophore crystal architectures}
Having established that denticle pulp cavities contain reflective iridophores and absorptive melanophores arranged differently across the three color zones (\cref{fig5}a), we next investigated how nanoscale crystal organisation within iridophores determines their optical behaviour. To address this, we combined microspectroscopic measurements of individual iridophores (\cref{fig5}b) with three-dimensional full-wave optical simulations based on 3D segmented crystal architectures (\cref{fig5}c,d). Although chromatophore distribution differs across zones, the spectral and angular response of individual iridophores is primarily governed by the internal orientation, spacing, and packing of their purine crystal arrays \citep{teyssier2015photonic, gur2024physical, funt2017koi}.\\
\input{Fig5}
Microspectroscopy of individual iridophores revealed distinct spectral signatures for blue-zone (BZ) and white-zone (WZ) denticles (\cref{fig5}b). Iridophores from the BZ exhibit a relatively narrow reflectance peak in the short-wavelength region centred in the violet–blue range, consistent with structural reflection from ordered multilayer purine stacks \citep{dearden2018sparkling, surapaneni2024ribbontail}. In contrast, iridophores from the WZ produce higher overall reflectance across the visible spectrum with a broader spectral profile characteristic of disordered crystal assemblies that scatter light more diffusely \citep{dearden2018sparkling, jordan2014disordered}.
To relate these optical signatures to nanoscale crystal organisation, we performed three-dimensional finite-difference time-domain (FDTD) simulations using crystal geometries segmented in 3D based on FIB–SEM of BZ and WZ iridophores (\cref{fig5}c,d). Transition-zone iridophores were not simulated because their crystal architecture closely resembles that of BZ iridophores, whereas their distinct optical appearance primarily arises from microscale chromatophore patterning. Simulations of the quasi-parallel crystal stacks reconstructed from BZ iridophores generated pronounced short-wavelength reflectance peaks with a maximum near \SI{480}{nm}, consistent with multilayer interference from ordered purine crystals \citep{pinsk2022biogenic, surapaneni2024ribbontail, blumer2024intermediate, zhang2023lizards, teyssier2015photonic}. By contrast, simulations based on the disordered crystal assemblies of WZ iridophores produced broader reflectance profiles indicative of incoherent scattering \citep{dearden2018sparkling, jordan2014disordered}.

Analysis of the far-field scattering patterns further highlights these differences. The reconstructed BZ architecture produces a strongly directional scattering distribution concentrated around the specular reflection direction, consistent with coherent reflection from ordered multilayer stacks. In contrast, the disordered crystal arrangement characteristic of WZ iridophores redistributes scattered light across a much broader angular range, generating diffuse broadband scattering.

These simulated optical responses closely match the experimentally measured spectral signatures of individual iridophores. Such contrasting optical behaviour is consistent with prior studies showing that quasi‑ordered multilayer purine arrays act as narrowband reflectors, whereas disordered purine crystal assemblies generate broadband reflectance or diffuse scattering in lizards, nudibranchs, and other vertebrates and invertebrates \citep{dearden2018sparkling, zhang2023lizards, funt2017koi}.

Notably, the simulated reflectance intensity of the reconstructed WZ crystal architecture is substantially lower than that of BZ stacks when considered in isolation. This difference reflects the intrinsic optical efficiency of the local crystal arrangements within individual iridophores. However, the measured reflectance represents the integrated optical response of the entire pulp‑cavity architecture, including the mesoscale organization of chromatophores. In BZ denticles, reflective iridophores are closely associated with absorptive melanophores that reduce the net reflected intensity, whereas WZ denticles lack melanophores and contain densely packed iridophores throughout the pulp cavity, a configuration reminiscent of chromatophore units in other fishes and cephalopods in which pigment cells modulate the apparent intensity and saturation of structural reflection \citep{surapaneni2024ribbontail, blumer2024intermediate, mathger2009mechanisms}. Consequently, microspectroscopic measurements integrate optical contributions across multiple spatial scales, explaining the higher overall reflectance observed in WZ denticles despite the lower intrinsic reflectivity of their individual crystal assemblies.

Together, the combined experimental and computational results demonstrate that nanoscale crystal architecture governs the fundamental optical response of shark iridophores, while the microscale organization of chromatophores modulates the intensity and spatial distribution of reflected light.

%% file: Fig1.tex
\begin{figure}[!ht]
\centering
\includegraphics[width=1\textwidth]{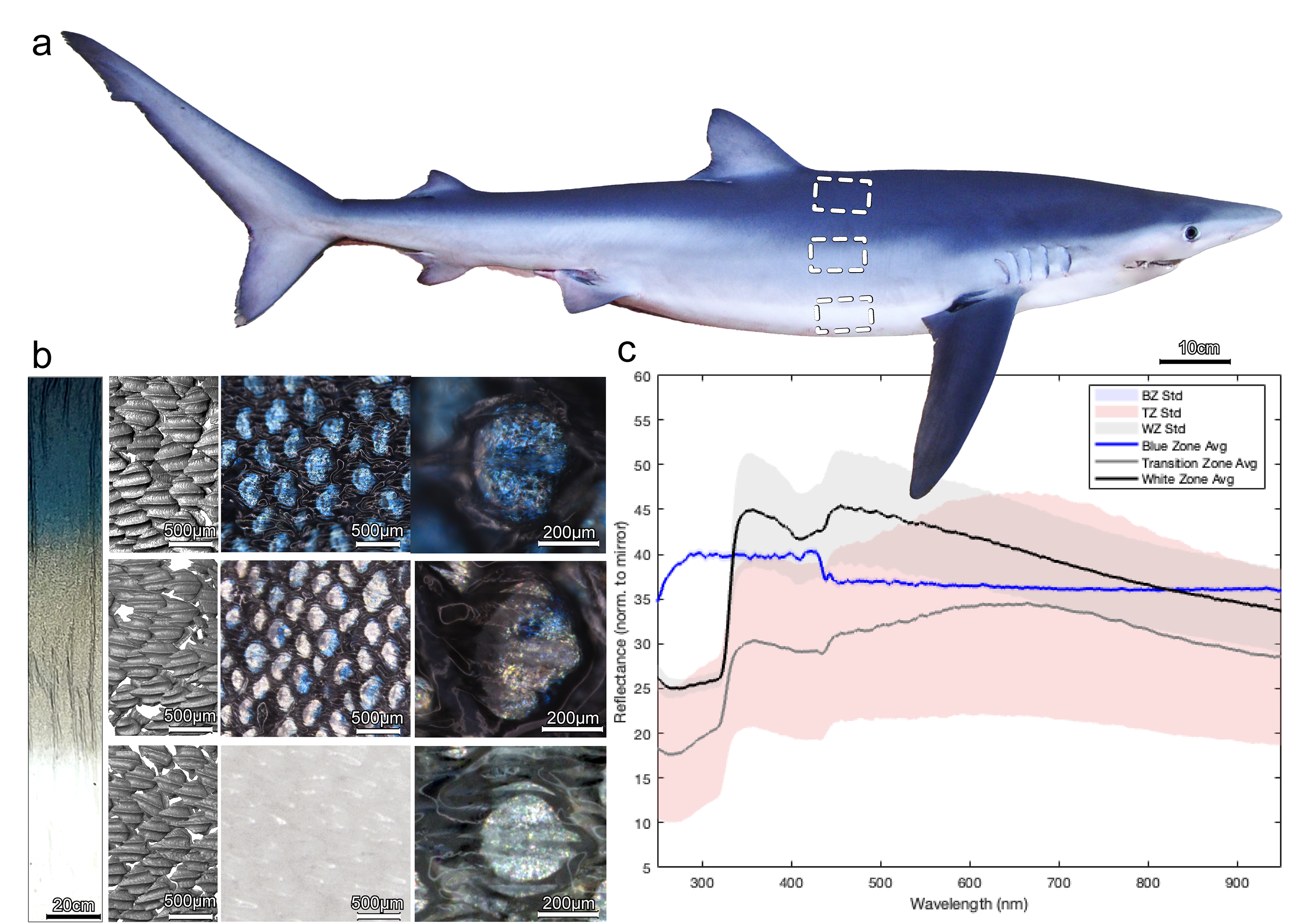}
\caption{\textbf{Structural and optical differences in denticles across blue shark skin color zones.}
(\textbf{a}) Photograph of a blue shark (image courtesy of Samuel Iglesias) showing the dorsal blue (\textbf{BZ}), lateral transition (\textbf{TZ}), and ventral white (\textbf{WZ}) regions used for sampling.
(\textbf{b}) Multimodal characterisation of denticles (apical views). (\textbf{Far left}) Optical image of a fresh skin strip showing the pronounced color gradient across zones. 
(\textbf{Left}) $\mu$CT reconstructions (\SI{2.5}{\micro\metre} resolution) revealing broadly similar external denticle morphologies across BZ, TZ, and WZ. 
(\textbf{Right}) Representative optical micrographs of denticles from each region showing distinct internal color domains. 
(\textbf{Far right}) High-magnification images of individual denticles confirm that color is directly linked to internal denticle architecture.
(\textbf{c}) Reflectance spectra collected from the three regions (n = \SI{9} per zone), with mean curves (\textbf{solid lines}) and standard deviations (\textbf{shaded areas}), highlighting distinct spectral signatures associated with each colour zone.}
\label{fig1_color_zones}
\end{figure}

%% file: Fig2.tex
\begin{figure}[htbp]
\centering
\includegraphics[width=0.93\textwidth]{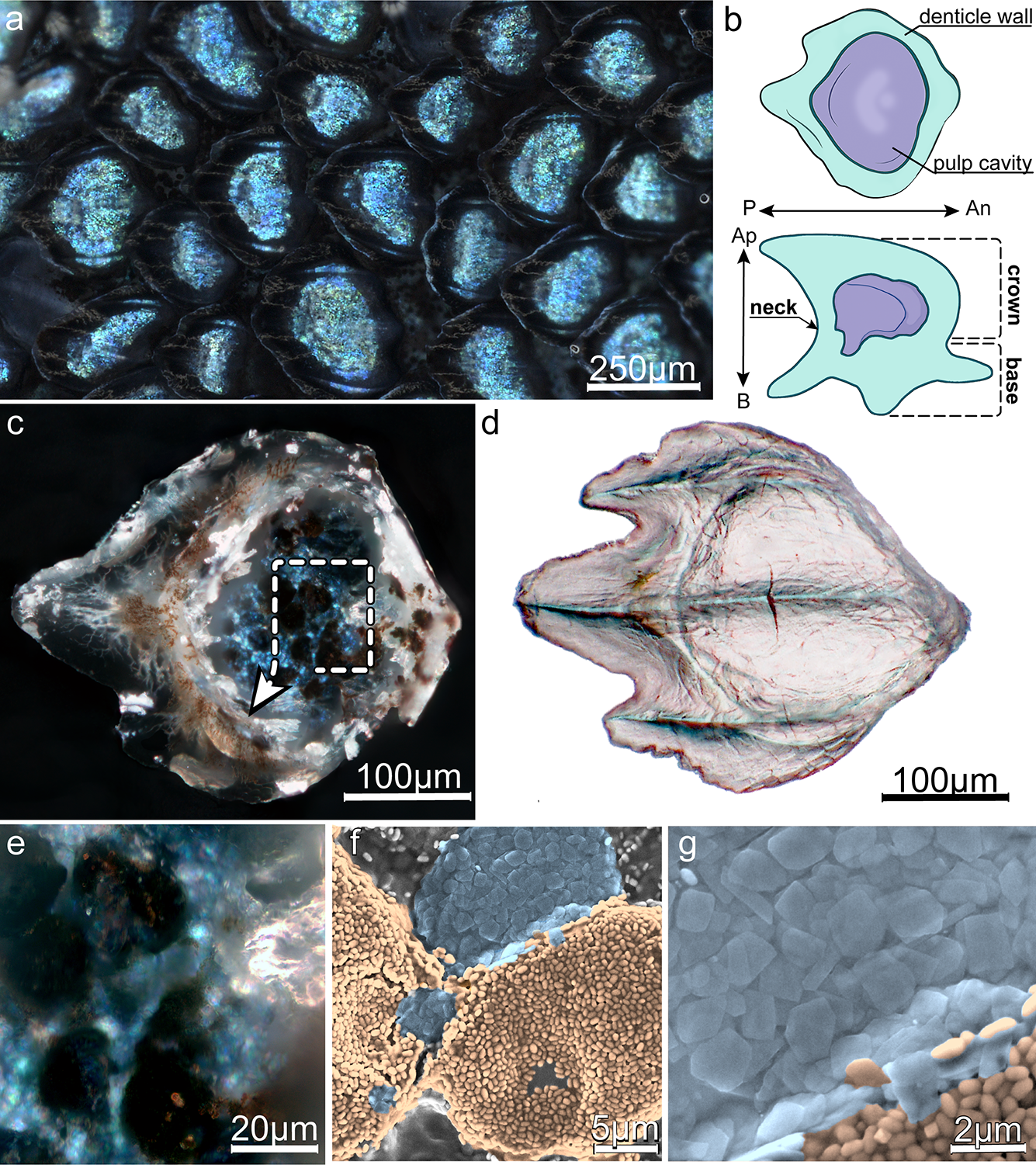}
\caption{\textbf{Origin of structural coloration: Nanoscale components within the blue shark denticle pulp cavity.}
(\textbf{a}) High-resolution optical micrograph of apically viewed denticles from the blue zone, oriented from right to left toward the shark’s tail, revealing distinct subsurface color domains within individual denticles.
(\textbf{b}) Schematic representation of a single denticle illustrating the position of the internal pulp cavity relative to the mineralized denticle wall. Top: apical view. Bottom: lateral view showing the crown and neck regions and the location of the pulp cavity within the denticle crown.
(\textbf{c}) Optical micrograph (z-stack, bright field and polarized light) of the basal side of a trimmed denticle crown (inverted orientation), exposing the pulp cavity; the dashed rectangle indicates the region shown at higher magnification in (\textbf{d}).
(\textbf{d}) Optical micrograph of the cleaned denticle crown (apical view) after centrifugation and ultrasonic treatment, demonstrating the transparency of the denticle wall once internal components are removed. (\textbf{e}) Higher-magnification optical micrograph of the}
\label{fig2_Structural_Comp}
\end{figure}

\begin{figure}[htbp]\ContinuedFloat
\caption{%
(cont'd.) 
boxed region in (\textbf{b}), revealing dark brown and bluish/shimmering internal domains within the pulp cavity.
(\textbf{f}) Environmental scanning electron micrograph (ESEM) of the same region shown in (\textbf{d}), with pseudo-coloring highlighting nanoscale structures associated with these optically distinct domains.
(\textbf{g}) Higher-magnification ESEM images showing two classes of nanoscale components within the pulp cavity: rod-like and spherical bodies, and flake-like platelets exhibiting variation in lateral dimensions and aspect ratios.
}
\end{figure}

%% file: Fig3.tex
\begin{figure}[htbp]
\centering
\includegraphics[width=0.98\textwidth]{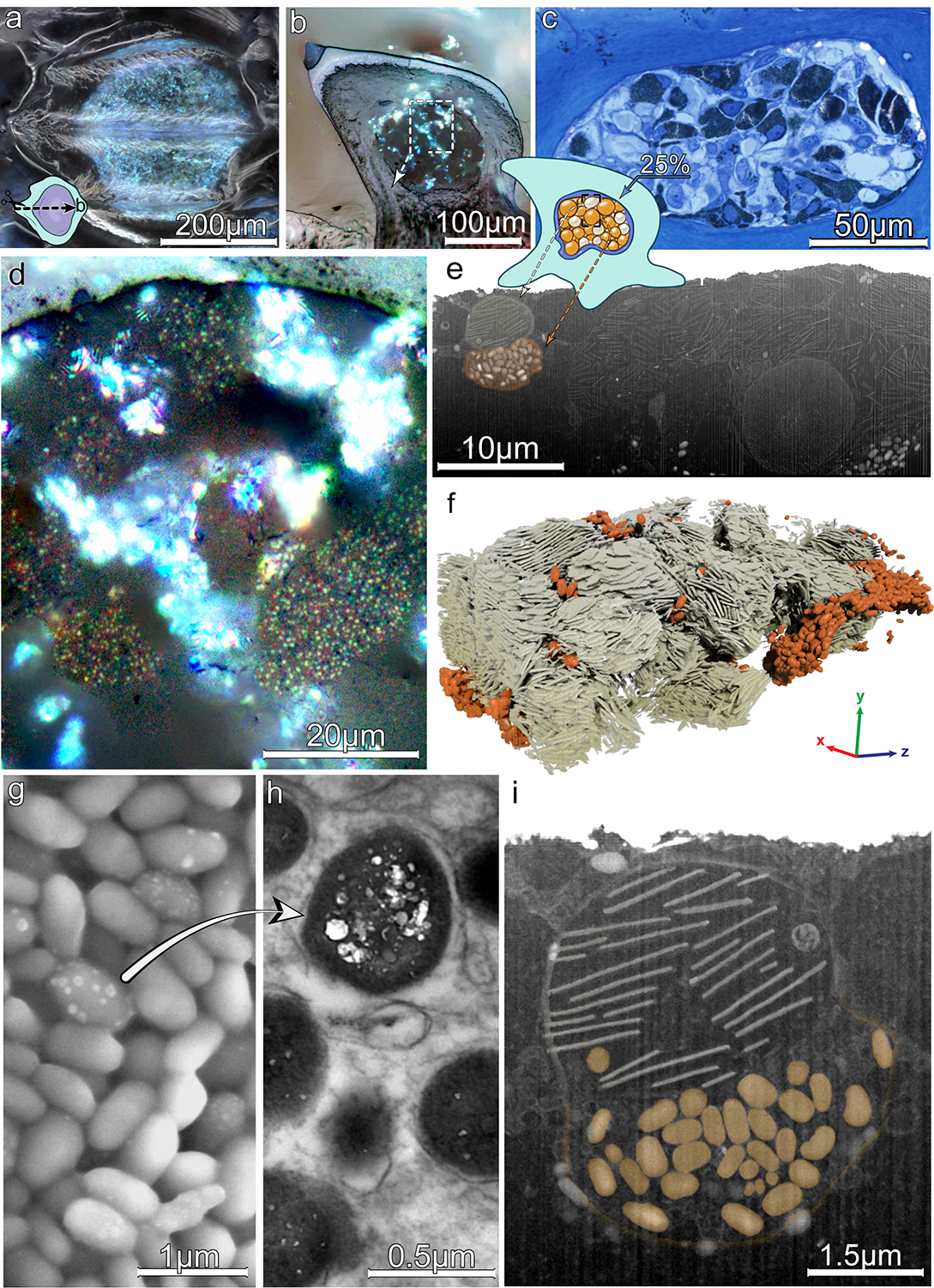}
\caption{\textbf{Micro- and nanoscale organization of color-producing components within blue-zone denticles.}
\textbf{(a)} High-resolution optical micrograph of denticles from the blue zone viewed apically, showing continuous blue coloration across the crown surface.
\textbf{(b)} Optical micrograph of a blue-zone denticle sectioned along
}
\label{fig3_BZ}
\end{figure}

\begin{figure}[htbp]\ContinuedFloat
\caption{%
(cont'd.)  the sagittal plane, revealing a crown-restricted pulp cavity occupying ~25\% of the total denticle volume. The cavity is densely packed with iridophores and melanophores arranged in a mixed, non-layered (“chessboard-like”) microscale organisation, as illustrated by the schematic overlay.
\textbf{(c)} Histological section of a blue-zone denticle showing the pulp cavity enclosed by the mineralized denticle wall.
\textbf{(d)} Higher-magnification optical micrograph resolving cellular and subcellular optical components.
\textbf{(e)} FIB–SEM cross-section beneath the denticle crown wall showing purine crystals within iridophores and melanosomes within melanophores.
\textbf{(f)} Three-dimensional FIB–SEM segmentation illustrating micro- and nanoscale organisation.
\textbf{(g)} ESEM of melanosomes revealing ($\sim$ \SIrange{10}{60}{nm}) protruding granules.
\textbf{(h)} TEM showing electron-dense internal inclusions.
\textbf{(i)} Higher magnification showing parallel-aligned purine crystal platelets.
Optical micrographs in \textbf{(b, d)} were generated from merged z-stacks using combined bright-field and polarized light microscopy.
}
\end{figure}

%% file: Fig4.tex
\begin{figure}[htbp]
\centering
\includegraphics[width=0.9\textwidth]{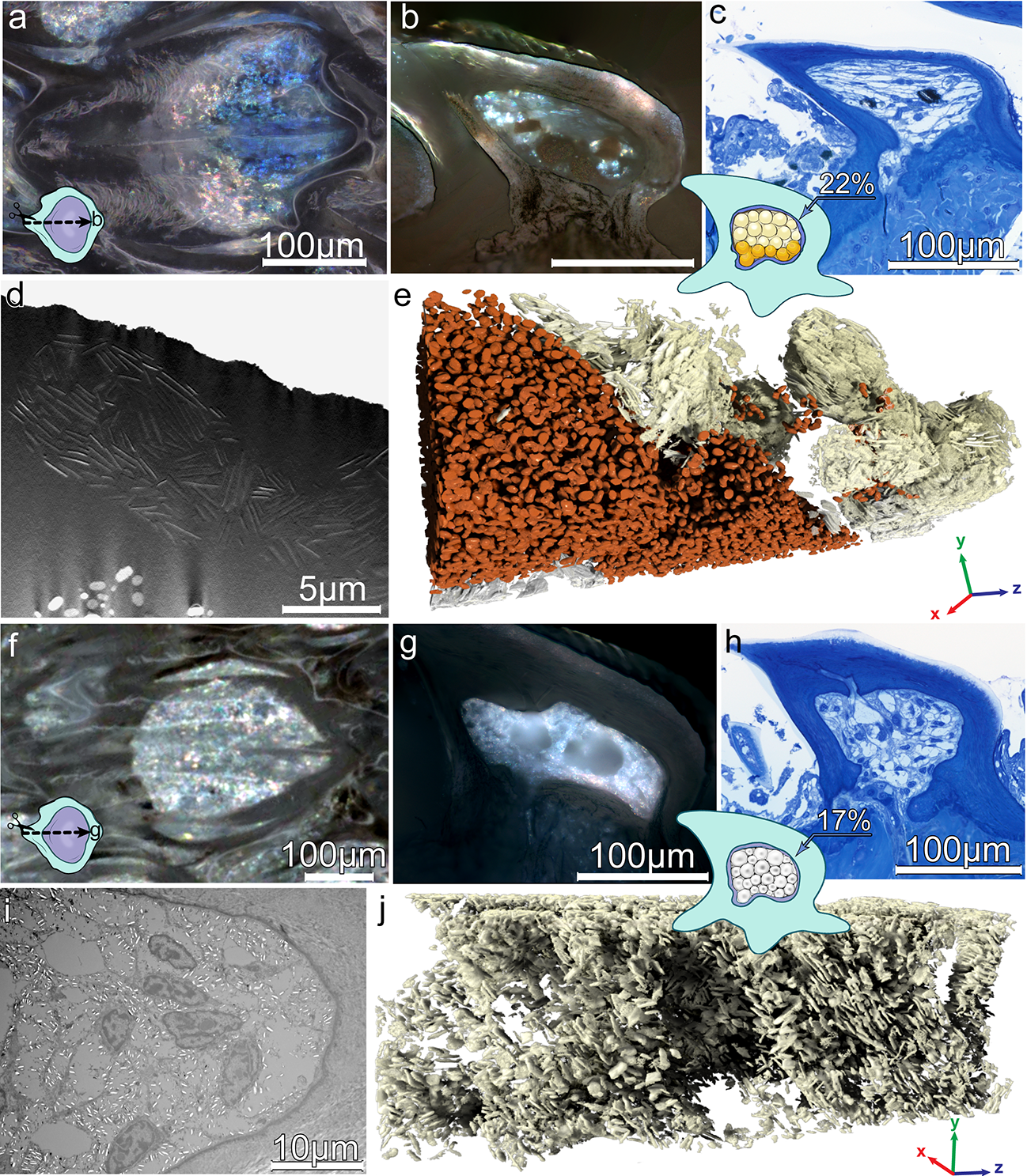}
\caption{\textbf{Micro- and nanoscale organization of color-producing components within transition- and white-zone denticles} \textbf{(a)} High-resolution optical micrograph of a denticle from the transition zone \textbf{(TZ)} viewed apically, showing mixed silver–blue reflectance at the crown surface.
\textbf{(b)} Optical micrograph of a \textbf{TZ} denticle sectioned along the sagittal plane, revealing a crown-restricted pulp cavity occupying ~22\% of the total denticle volume. The cavity is densely packed with iridophores and melanophores arranged in a layered microscale pattern, as illustrated by the schematic overlay. \textbf{(c)} Histological section of a transition-zone denticle showing the crown-restricted pulp cavity enclosed by the mineralized denticle wall and connected to the underlying dermis. Darkly stained bodies correspond to melanophores within}
\label{fig4_TZ_WZ}
\end{figure}

\captionsetup{type=figure}
\addtocounter{figure}{-1}
\captionof{figure}{%
(cont'd). the pulp cavity, whereas unstained regions occupy positions that spatially coincide with the reflective domains observed in optical micrographs \textbf{(b)}.
\textbf{(d)} place holder
\textbf{(e)} Three-dimensional segmentation of FIB–SEM data from a transition-zone denticle illustrating the coupled micro- and nanoscale organisation of color-producing components: reflective purine crystals (ivory) confined within iridophores and absorptive melanosomes (orange) within melanophores, arranged in a layered microscale pattern.
\textbf{(f)} High-resolution optical micrograph of a denticle from the white zone \textbf{(WZ)} viewed apically, showing bright, broadband white reflectance across the crown surface.
\textbf{(g)} Optical micrograph of a \textbf{WZ} denticle sectioned along the sagittal plane, revealing a crown-restricted pulp cavity occupying ~17\% of the total denticle volume. The cavity is densely filled with shimmering reflective components, as illustrated by the schematic overlay, with no apparent absorptive elements. \textbf{(h)} Histological section of a white-zone denticle showing a crown-restricted pulp cavity enclosed by the mineralized denticle wall and connected to the underlying dermis. The cavity contains unstained regions corresponding to the locations of purine crystal–rich domains observed in optical and electron microscopy, while absorptive components are absent.
\textbf{(i)} Transmission electron micrograph \textbf{(TEM)} of the \textbf{WZ} pulp cavity showing irregularly shaped iridophore-like compartments densely filled with purine crystals.
\textbf{(j)} Three-dimensional segmentation of FIB–SEM data from a \textbf{WZ} denticle showing that purine crystals are not organized into discrete, spheroidal iridophores but instead form irregular, continuously wrapped membrane-bound structures containing chaotically arranged crystals.}
\vspace{2\baselineskip}

%% file: Fig4_morph.tex
\begin{figure}[htbp]
\centering
\includegraphics[width=0.85\textwidth]{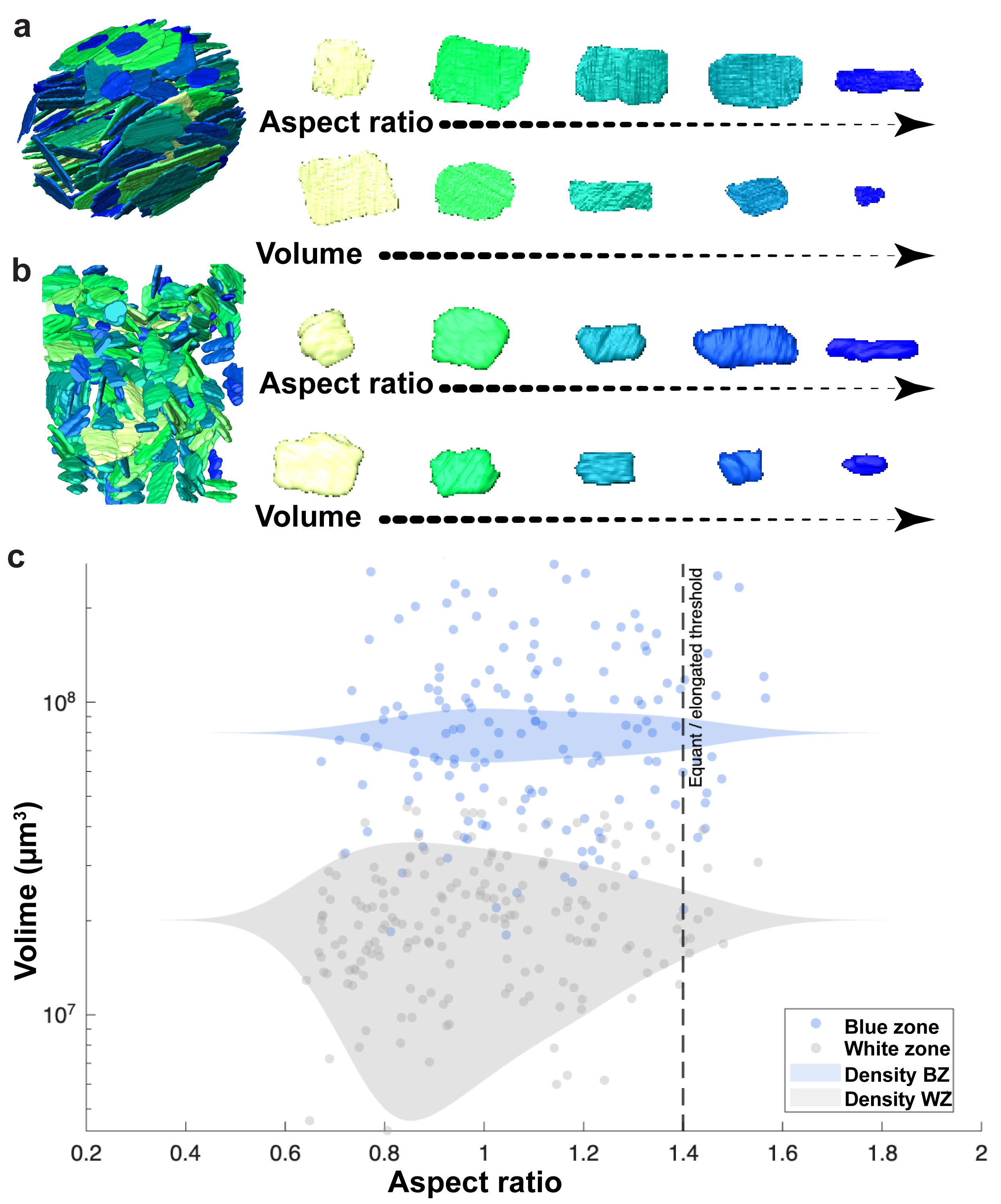}
\caption{\textbf{Morphological diversity and morphospace distribution of guanine crystals} \textbf{(a)} Representative guanine crystals from the blue zone \textbf{(BZ)} illustrating morphological variability across the dataset. Crystals are arranged to reflect increasing elongation and volume. \textbf{(b)} Representative guanine crystals from the white zone \textbf{(WZ)}, showing the range of observed crystal morphologies in this region. \textbf{(c)} Morphospace distribution of individual crystals from the \textbf{BZ} (blue) and \textbf{WZ} (grey) plotted as crystal volume versus aspect ratio. Each point represents a single crystal. Diamond markers indicate the median values for each zone. The dashed vertical line denotes the equant–elongated threshold (aspect ratio = 1.4). Crystal volume is shown on a logarithmic scale. n = 152 crystals from the blue zone and n = 137 crystals from the white zone.}
\label{fig4_morph}
\end{figure}

%% file: Fig4_angle.tex
\begin{figure}[htbp]
\centering
\includegraphics[width=1\textwidth]{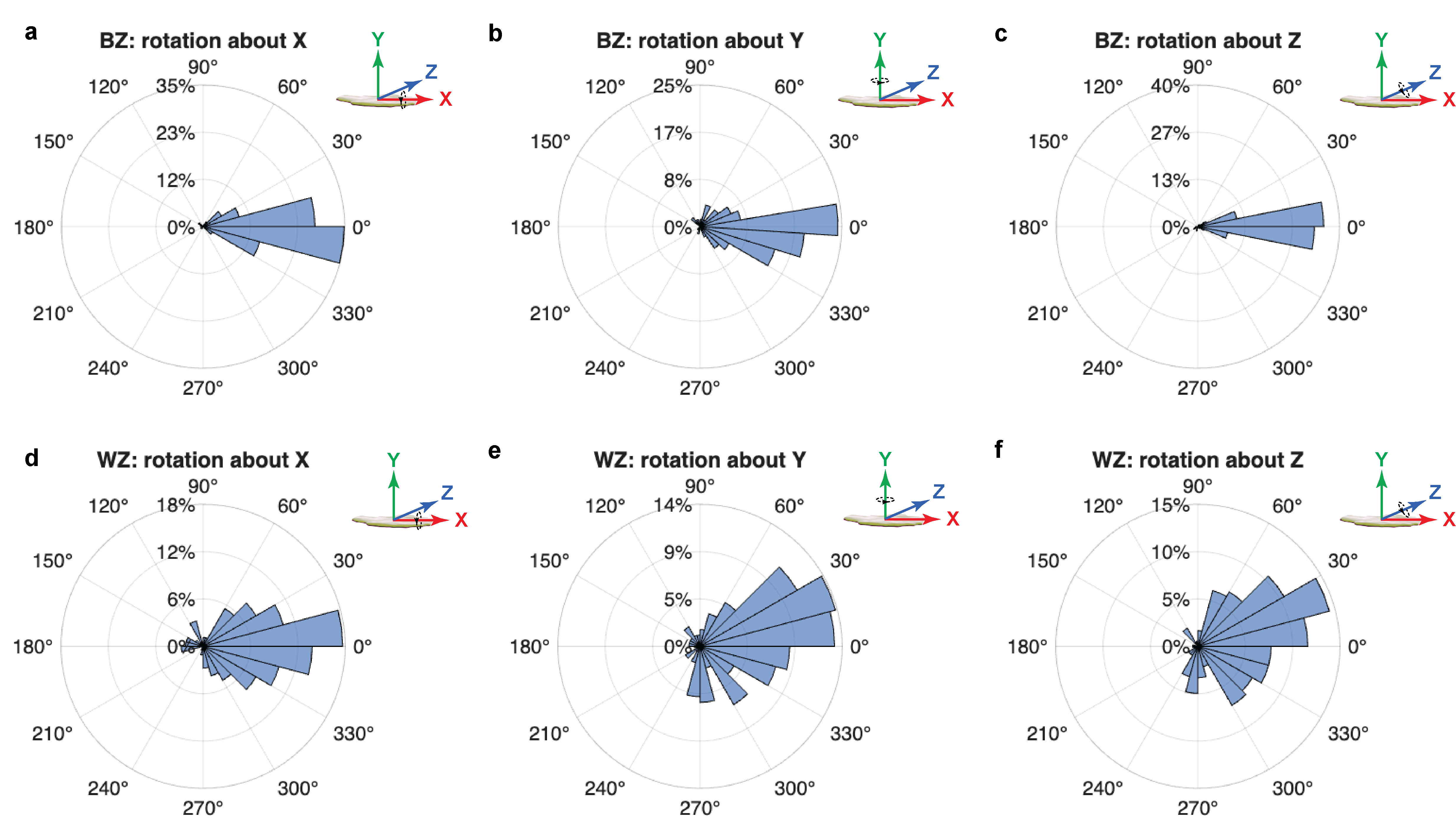}
\caption{\textbf{Rose diagrams of purine crystal orientation in blue- and white-zone iridophores.} Circular histograms showing angular distributions of individually segmented purine crystals relative to the denticle reflecting surface. Crystal orientation was defined by the principal eigenvector of each segmented crystal. \textbf{a–c}, Blue-zone \textbf{(BZ)} iridophores (n = 152 crystals). \textbf{d–f}, White-zone \textbf{(WZ)} iridophores (n = 137 crystals). Rotations are shown about the X \textbf{(a,d)}, Y \textbf{(b,e)}, and Z \textbf{(c,f)} axes.}
\label{fig4_angle}
\end{figure}

%% file: Fig5.tex
\begin{figure}[htbp]
\centering
\includegraphics[width=0.9\textwidth]{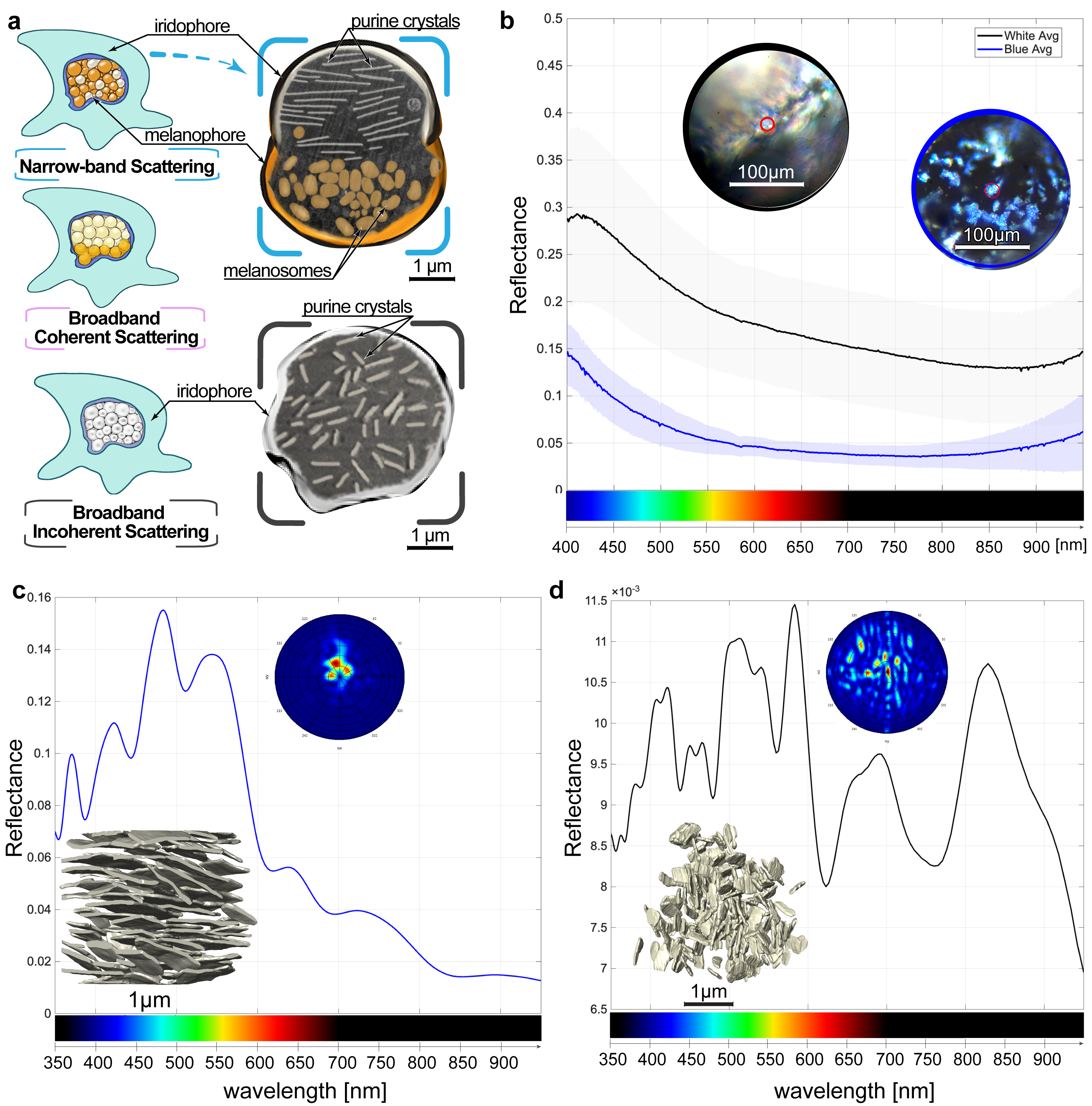}
\caption{\textbf{Optical response of iridophores from microspectroscopy and optical modelling.} \textbf{(a)} Conceptual model of multiscale color production in denticle pulp cavities. Sagittal sections show cavities densely packed with reflective iridophores (bright) and absorptive melanophores (dark). Their three-dimensional organization differs across color zones, producing mixed arrangements in the blue zone \textbf{(BZ)}, layered organization in the transition zone \textbf{(TZ)}, and melanophore-free iridophore assemblies in the white zone \textbf{(WZ)}. At the nanoscale (right), optical behavior is governed by the three-dimensional architecture of guanine crystals within iridophores. \textbf{(b)} Microspectroscopic reflectance spectra from individual iridophores in BZ (blue) and WZ (black) denticles. BZ iridophores exhibit a narrow short-wavelength reflectance peak, whereas WZ iridophores show higher overall reflectance and broader spectral scattering. (c, d) Three-dimensional optical simulations of light reflection from guanine crystal architectures reconstructed from FIB–SEM segmentations of iridophores from BZ \textbf{(c)} and WZ \textbf{(d)}. Simulations were performed using finite-difference time-domain (FDTD) modelling, with purine crystals (n = 1.7) embedded in a cytoplasmic medium (n = 1.33) and illuminated by a broadband plane-wave source (300–800 nm). Ordered crystal stacks characteristic of BZ iridophores produce directional short-wavelength reflectance, whereas the more disordered crystal assemblies of WZ iridophores generate broader scattering spectra.}
\label{fig5}
\end{figure}

%% file: 3.Conclusion.tex
\subsection*{Conclusion}
Although the blue shark’s striking dorsoventral color gradient is widely known and interpreted as pelagic countershading, its physical basis has remained unresolved, particularly how the skin generates a spectrally narrowband blue dorsum while producing broadband silver and white ventrally, and what role dermal denticles—best known for their hydrodynamic functions—play in this optical system. Our results reveal a multiscale photonic architecture in which a minimal set of optical components—reflective purine crystals and absorptive melanosomes—are arranged within denticle pulp cavities to generate the blue, silver, and white appearances observed across the shark’s body surface.

By combining ultrastructural imaging, microspectroscopy, and optical simulations, we show that the shark’s color gradient arises from hierarchical organization within denticle pulp cavities, where microscale chromatophore arrangements and nanoscale purine crystal architectures inside iridophores together control reflected color across the shark’s body surface.

Across the body surface, these architectures form distinct optical configurations: dorsal denticles contain quasi-parallel purine crystal stacks extending across iridophores and coupled with melanophores that generate narrowband blue reflection; transition-zone denticles retain quasi-parallel crystals but in shorter, slightly misaligned stacks within a layered chromatophore arrangement that produces a silvery appearance; and ventral denticles lack melanophores and contain highly disordered crystal assemblies that generate broadband white scattering.

Comparable purine-based photonic architectures occur across diverse biological systems, including vertebrate iridophores and reflective tissues in invertebrates, where the organization of guanine platelets controls reflected color. By revealing that such architectures are localized within individual dermal denticles, our findings uncover an unexpected optical function for structures traditionally associated with hydrodynamics and establish blue shark denticles as mechanically protected optical “pixels”, whose coupled microscale chromatophore organization and nanoscale purine crystal architectures generate the shark’s countershaded color gradient

%% file: 4.Experimental_Section.tex
\newpage
\section*{Experimental Section}
\subsection*{Sample Collection and Handling:}
Fresh specimens of the blue shark (\textit{Prionace glauca}, Linnaeus 1758) were obtained from commercial fisheries in the north-eastern Atlantic Ocean (Bay of Biscay and Brittany, France) and the western Pacific (Taitung County, Taiwan). Capture location, date, sex, and morphometrics were recorded following the established protocol by \cite{mollen2019making}.\\
Specimens analysed in this study included: an adult female (142 cm total length) obtained from Chenggong fish market, Taitung County, Taiwan (March 2024); adult females ERB 1350 (\SI{161.0}{cm}) and ERB 1351 (\SI{187.0}{cm}) obtained from Roscoff fish market, Brittany, France (July 2024); and adult female ERB 1376 (\SI{181.0}{cm}) and juvenile male ERB 1377 (\SI{121.3}{cm}) obtained from Ondarroa, Spain (Bay of Biscay, FAO 27.8b; July 2025).\\
Micro- and nanoscale structural analyses were primarily conducted on specimens ERB 1350, ERB 1376, and ERB 1377.\\
All individuals originated from legal commercial fisheries and were sampled post-mortem. The blue shark has been listed under CITES Appendix II since 2023. For European specimens, required certificates of origin were obtained, and transport complied with EU and CITES regulations.
The Taiwanese specimen was obtained from a legally operating fish market during the open fishing season in collaboration with local marine researchers. All analyses were conducted in Taiwan, and no biological material was exported.
As no live animals were handled or euthanized for this study, ethical approval was not required.

Skin samples were collected from defined body regions (dorsal, lateral and ventral), encompassing blue, transitional and white color zones. Specimens and regions were allocated to specific analytical pipelines according to methodological requirements, as detailed in the corresponding subsections below.

\subsection*{Light microscopy (bright-field and polarized light)}
Bright-field (BF) and polarized light (PL) microscopy were performed in reflection mode using a Zeiss Axiolab 5 (Carl Zeiss GmbH). PL images were acquired with both the polarizer and analyser inserted using a Nikon Plan 100X/0.90 (working distance \SI{0.26}{mm}). Sagittal sections of denticles from the three color zones (BZ, TZ, WZ) were imaged as Z-stacks at identical magnification (50×) using both BF and PL. The pulp cavity was imaged from the basal side of trimmed denticle crowns at 50× and at higher magnification (100×) under polarized light. Exposure times ranged from 20–100 ms depending on imaging mode. Z-stacks were combined in Adobe Photoshop (version 2025) using the Auto-Blend Layers function to generate extended-depth-of-field images.

\subsection*{Macroscopic reflectance spectroscopy and stereomicroscopy}
Optical measurements were conducted on freshly obtained ex vivo skin from the Taiwanese specimen described in the Specimen Collection and Handling section. Measurements were performed on unfixed tissue immediately upon arrival at the laboratory facility.\\
Dorsal--ventral skin regions posterior to the right pectoral fin were examined, encompassing blue, transitional, and white color zones. Tissue samples were kept moist with seawater and positioned flat on a non-reflective surface. All measurements were conducted in a dark room to eliminate interference from ambient light.\\
Macroscopic reflectance spectra were acquired using a Miniature Spectrometer OceanHDX (Ocean Insight) coupled to a UV--VIS--NIR light source (DH-2000-BAL, Ocean Insight). Illumination and collection were achieved using a bifurcated reflectance probe (fiber type: UV--VIS; core diameter \SI{300}{\micro\meter}; length \SI{2}{m}; part \#TP300-UV-VIS; assembly \#EOS-A4000120; Ocean Insight) with an approximate sampling spot diameter of \SI{3}{mm}. The probe was mounted in a custom-designed transparent holder that maintained perpendicular alignment to the skin surface and a fixed working distance of \SI{2}{mm} from the denticle layer to ensure consistent measurement geometry.\\
For each of the three color regions, spectra were recorded from three spatially distinct spots, with three technical replicates acquired per spot. Data acquisition was performed using OceanView software (version 2.0.1.4). Subsequent spectral processing was conducted in MATLAB (R2024a, MathWorks), where measurements were averaged per color region, and standard deviations were calculated to quantify variability.\\
Immediately following spectral acquisition, the identical measurement locations were documented using a Nikon SMZ18 stereomicroscope (Nikon Corporation, Tokyo, Japan) to correlate macroscopic optical output with surface denticle morphology.

\subsection*{Digital microscopy and optical micro-spectroscopy}
Low magnification images were obtained using a Keyence VHX-7100 digital microscope (Keyence Corporation, Japan) equipped with E20, E100, and E500 lenses. Optical imaging and micro-spectroscopy were performed on individual scales using a customized Zeiss AxioScope 7 equipped with a water immersion 40x objective (Zeiss, WN-Achroplan, NA 0.75) and using a halogen lamp as light source (HAL100, Zeiss). Reflectance spectra were recorded on an area of \SI{17.4}{\micro\meter} in diameter by coupling the microscope to a spectrometer (Avantes, HS2048XL-EVO) via an optical fiber (Avantes, FC-UVIR600-2, \SI{600}{\micro\meter} core size) and referenced against a protected silver mirror (Thorlabs, PF10-03-P01). Micrographs were acquired with a high frame rate camera (Pixelink, 	PL-D7620CU-T), calibrated against a white standard (Avantes, WS-2). Artificial seawater (\SI{33}{ppt}, Instant Ocean sea salts) was used as the medium between the water immersion objective and the scales.  

\subsection*{Histology and transmission electron microscopy}
Skin samples used for histological and ultrastructural analyses were derived from specimen ERB~1376. Three dorso--ventral skin strips (ca. \SI{1.5}{cm} $\times$ \SI{30.0}{cm} $\times$ \SI{1.5}{cm}) were excised and immersion-fixed immediately post-mortem in \SI{4}{\%} paraformaldehyde (PFA) directly on the fishing vessel deck, following the natural death of the animal during standard fishing operations. Within \SI{24}{h}, the strips were subdivided into smaller blocks (ca. \SI{0.5}{cm^3}) representing dorsal, lateral and ventral regions, each encompassing blue, transitional and white color zones. Subsamples were maintained either in PFA or transferred to paraformaldehyde--glutaraldehyde (Karnovsky) fixative and stored at approximately \SI{6}{\degreeCelsius} prior to processing.

Tissue was rinsed in \SI{0.1}{M} PBS and post-fixed overnight at \SI{4}{\degreeCelsius} in \SI{0.5}{\%} osmium tetroxide and \SI{1}{\%} potassium ferricyanide in distilled water. Following rinsing, samples were demineralized for \SI{3}{d} at \SI{37}{\degreeCelsius} in EDTA (ethylenediaminetetraacetic acid), dehydrated through a graded ethanol series and subsequently in acetone, and embedded in EPON resin (\#45359, Sigma-Aldrich, Austria).

Vertical serial semi-thin sections (\SI{1}{\micro\meter}) were cut using a Reichert Ultracut~S microtome (Leica Microsystems, Wetzlar, Germany) equipped with a Histo-Jumbo diamond knife (Diatome, Biel, Switzerland). Sections were stained with toluidine blue at \SI{60}{\degreeCelsius} for \SI{3}{min} and examined by bright-field microscopy to guide subsequent transmission electron microscopy (TEM) analysis of ultrathin sections from the same resin blocks. Semi-thin sections were imaged using a Zeiss Axio10 microscope equipped with an AxioCam~512 color digital camera and ZEN~3.0 Blue Edition software (Zeiss, Oberkochen, Germany).

Vertical serial ultra-thin sections (\SI{90}{nm}) were cut using an ultra-diamond knife (Diatome, Biel, Switzerland), mounted on dioxane--formvar-coated slot grids (\#PYSL2010S-CU, Science Services, Munich, Germany), and examined without post-staining. TEM imaging was performed at \SI{80}{kV} using a Philips CM120 transmission electron microscope (FEI, Eindhoven, The Netherlands) equipped with a MORADA digital camera and iTEM software (Olympus SIS, Münster, Germany).

\subsection*{Micro-Computed Tomography}
Micro–computed tomography (µCT) imaging was performed using a Comet Yxlon FF35 CT scanner (Comet Yxlon, Hamburg, Germany) at The Hong Kong Polytechnic University. Specimens were mounted on a low-attenuation carbon fibre rod to minimise background artefacts during scanning. The X-ray source was operated in transmission mode at \SI{70}{kV} and \SI{95}{\micro A}. A total of 3600 projections were acquired over a \SI{360}{\degree} rotation. No X-ray filter was applied due to the low density of the specimen.\\
Tomographic reconstructions were performed using Yxlon Reconstruction Workspace (version 2206.4.0) without additional correction algorithms, yielding isotropic voxel resolutions of \SI{2.17}{\micro\meter} and \SI{2.00}{\micro\meter} for the individual scans.

\subsection*{Environmental Scanning Electron Microscopy (ESEM), wet mode}
Fresh individual denticles were placed on conductive carbon tape mounted on aluminum SEM stubs. Secondary electron imaging was completed under environmental (wet) conditions in an SEM instrument operated at 8--10kV (Quattro S, Thermo Fisher Scientific). The instrument is equipped with a gaseous secondary electron detector and a Peltier cooling stage. Relative humidity (RH) within the chamber was controlled by adjusting chamber pressure while maintaining the Peltier cooling stage at 3--4 °C. Experiments were performed from low (20-30\% RH) to high (100\%) humidity conditions in 10-20\% increments, or conversely 

\subsection*{Focus Ion Beam Scanning Electron Microscopy (FIB-SEM)}
\subsubsection*{FIB-SEM for the sample from Blue zone}
Resin blocks containing tissue samples were polished in order to expose the tissue at the block surface. Samples were sputter-coated with a conductive layer is 10-20 nm Pt layer and transferred to the Zeiss Crossbeam 550 (Carl Zeiss Microscopy GmbH, Germany). Serial surface imaging was performed using the ATLAS 3D: nanotomography software package (A3D) of the ATLAS engine v. 5.3.5.27 (Fibics, Ottawa, Canada). The A3D nanotomography run was conducted as follows. First, the region of interest (40 $\mu$m wide and 25 $\mu$m high) was prepared by depositing a protective platinum layer. Then, 3D tracking and autotune fiducial marks were milled to perform automated tracking and autofocus. Milled fiducial marks were highlighted with a deposited carbon layer, and subsequently, a carbon protective layer was applied to the entire preparation pad. High FIB probe currents (\SI{30}{kV}; \SI{30}{nA} and \SI{3}{nA}) were used to create a trench in the specimen to expose the initial 3D imaging surface. Automatic serial sectioning and imaging were then performed using \SI{700}{pA} as imaging and milling probe currents. The accelerating voltage for the electron probe was set to \SI{2.5}{kV}. Interlaced volumetric imaging was performed, setting a \SI{10}{nm} isotropic voxel size. Signals from the secondary electron detector (Inlets) and the energy-selective backscattered electron (EsB, grid voltage set to \SI{1.5}{kV} detector were collected.
\subsubsection*{FIB-SEM for the sample from Transition and White zone}
The PMMA-embedded white (WZ) and transition zone (TZ) denticles were mounted on aluminum SEM stubs using conductive carbon adhesive tabs. Samples were coated with a 10 nm carbon layer followed by a 10 nm platinum layer using a Safematic CCU-010 high-vacuum coating system. To further enhance conductivity, carbon stripes were applied around the mounted specimens.
FIB-SEM acquisition was performed using the crossbeam 540 (Zeiss, Oberkochen). A trapezoidal trench was milled to expose a cross-sectional imaging surface, with approximate dimensions of 60 $\mu$m at the bottom length, 50 $\mu$m at the top, and 35 $\mu$m in height. Initial coarse milling was carried out at 30 kV. For serial sectioning during volume acquisition, the ion beam current was reduced to either 1.5 nA (TZ) or 700 pA (WZ). Serial imaging was performed with the electron beam operated at 2 kV and a probe current of 600-700 pA. Images were acquired in dual-detector mode using a mixed signal from the InLens and secondary electron (SE) detectors, as well as the backscattered electron (BSE) detector. The image resolution was set to 4096 × 3072 pixels. A voxel size of 10 nm was used in the x, y, and z directions to obtain isotropic datasets.
\subsection*{Three-dimensional segmentation}
\subsubsection*{Three-dimensional segmentation of denticles}
Three-dimensional segmentation of denticles and their internal pulp cavities was performed on reconstructed micro–computed tomography (µCT) datasets to quantify the relative volume of mineralized denticle tissue and enclosed cavity space. Segmentation was conducted in Amira (ZIB Edition, v. 2022 Zuse Institute, Berlin) using a semi-automated workflow. An intensity-based threshold was applied to isolate the high-attenuation signal corresponding to the mineralized denticle wall (enamel and dentine). Individual denticles were then separated and assigned to independent label fields. The internal pulp cavity was segmented by filling the enclosed space within each denticle wall, producing a separate cavity label.
Segmentation was performed for ten denticles from each skin color zone. Three-dimensional objects were generated, and the volumes of the mineralized denticle wall and the associated pulp cavity were measured using the object analysis module in Amira. The proportion of denticle volume occupied by the pulp cavity was then calculated.

\subsubsection*{Three-dimensional segmentation of cellular and intracellular organizations}
Three-dimensional segmentation of FIB–SEM datasets was performed to enable both qualitative visualization and quantitative analysis of iridophore architecture. Qualitative exploration of cellular organization and intracellular compartmentalization within the FIB–SEM volumes was conducted using the AI-based segmentation module in Dragonfly, exclusively for three-dimensional rendering and exploratory visualization. For quantitative analysis, individual guanine crystals within selected iridophores were segmented in Amira as separate three-dimensional objects. This object-level segmentation enabled the extraction of crystal-specific geometric and spatial descriptors, including volume, aspect ratio, principal orientation vectors, and center-of-mass coordinates. These parameters served as the basis for all subsequent analyses of crystal morphology, angular distributions, and inter-crystal spacing in both Blue Zone (BZ) and White Zone (WZ) iridophores and also as inputs for computational optical simulations.
\paragraph{Segmentation of Blue Zone (BZ) iridophores.}
For the Blue Zone iridophore, based on the FIB–SEM volumetric dataset, individual guanine crystals inside a cellular-shaped iridophore were segmented in three dimensions using Amira software (ZIB Edition, v. 2022 Zuse Institute, Berlin). Segmentation followed a semi-automated workflow combining grayscale-based thresholding with manual refinement. Crystal surfaces were initially labeled on two-dimensional XY slices at regular intervals, leaving 3–5 slices unsegmented between annotated planes. Then, missing labels were reconstructed using interpolation, which reduced the influence of FIB-induced image-jump artifacts that can distort three-dimensional models.

\paragraph{Segmentation of White Zone (WZ) iridophores}
For the White Zone iridophore, three-dimensional segmentation of guanine crystals was performed on the FIB–SEM volumetric dataset using the AI-based segmentation module in Amira software (Amira3D for EM, 2024.2 Edition, Trial Thermo Scientific License). In contrast to the BZ iridophore, the WZ iridophore exhibited an irregular and extended cellular morphology, with no well-defined spherical boundaries, and contained guanine crystals with highly variable and isotropic orientations. AI-based segmentation was trained across the full FIB–SEM volume using manual labels of guanine crystals on five representative 2D slices, which served as training references for the neural network. To improve segmentation accuracy, five iterative post-segmentation training cycles were performed. These cycles were used to progressively reduce segmentation artifacts and to recover crystals missed in earlier iterations, particularly those with orientations or contrast levels that challenged the initial model. To enable direct comparison with the BZ dataset, a subvolume containing a comparable number of crystals was cropped from the WZ iridophore. The resulting segmentation was subsequently filtered to remove isolated artifacts and non-crystalline features. Unlike the BZ workflow, in which crystals were segmented directly as individual objects, initial WZ segmentation produced clusters of crystal belong to one singe label. A morphological separation module was applied to divide disconnected three-dimensional objects into individual labels. Residual cases in which adjacent crystals remained partially connected by a small number of voxels were resolved by targeted manual separation to ensure that each crystal was represented as an independent three-dimensional object.

\subsubsection*{Morphological analysis of guanine crystals}
Following three-dimensional segmentation, each guanine crystal was represented as an independent labeled object. Morphological descriptors were extracted using the label analysis module in Amira (Amira software (ZIB Edition, v. 2022 Zuse Institute, Berlin). For each crystal, geometric parameters including volume and aspect ratio were computed automatically based on the three-dimensional voxel representation of the segmented objects. To visualize morphological diversity within the datasets, crystals were arranged using the Grid module in Amira, which allowed sorting of individual objects according to selected geometric descriptors. Crystals were ordered by increasing aspect ratio and volume, and color-coded based on the corresponding quantitative values obtained from the label analysis. Morphospace distributions were generated by plotting crystal volume against aspect ratio for individual crystals segmented from Blue Zone (BZ) and White Zone (WZ) iridophores. Each point in the resulting distribution represents a single crystal extracted from the FIB–SEM dataset. Median values for each region were calculated to facilitate comparison of characteristic crystal morphologies between zones.
\subsubsection*{Quantification of angular distributions}
Crystal orientation was quantified by extracting the principal orientation vector for each segmented guanine crystal, defined as the principal eigenvector corresponding to the crystal’s longest geometric axis. Angular deviations were calculated relative to the laboratory coordinate system (X, Y, Z), enabling characterization of crystal tilt and rotational dispersion within the iridophore volume. Orientation data were analyzed separately for the Blue Zone (BZ) and White Zone (WZ) datasets. To remove offsets introduced by global sample orientation during FIB–SEM acquisition, angular values were centered by subtracting the population means for each axis. This normalization isolates intrinsic orientational dispersion within each crystal population. Angular distributions were visualized using circular histograms (rose diagrams) showing the frequency distribution of crystal orientations around each rotational axis. Separate plots were generated for the X, Y, and Z rotations for both BZ and WZ datasets. All analyses and visualizations were performed in MATLAB (MathWorks R2025b).unbiased comparison between regions.

\subsection*{Optical modelling}
Three-dimensional electromagnetic simulations were performed using Lumerical FDTD Solutions 2025 R2 (Ansys Lumerical) to calculate the angle-dependent reflectivity of the reconstructed intracellular architecture. The simulation geometry was directly generated from the three-dimensional tomographic reconstruction and imported into the 3D FDTD simulation environment. Regions corresponding to the reconstructed crystalline material were assigned the refractive index of 1.7 (purine), while the surrounding medium was modelled as a homogeneous cytoplasmic aqueous matrix with a refractive index of 1.33, that of water.

The structure was illuminated with a normally incident broadband plane-wave source spanning 350--950 nm. The optical response of the system was then measured with a reflectivity monitor placed above the light source and the structure. Perfectly matched layer (PML) boundary conditions were applied on all boundaries of the simulation domain to suppress non-physical reflections from the domain edges. Reflectance spectra were obtained by monitoring the backward-propagating electromagnetic fields and calculating the wavelength-dependent reflected power relative to the incident power, as well as the corresponding far-field scattering patterns. All simulations were conducted in 3D to capture the effects of the complex tomographically resolved geometry on the measured reflectance.
\newpage